\documentstyle[aps]{revtex}

\begin{document}

\begin{center}
{\bf Edge magnetoplasmons in periodically modulated structures}

\bigskip

O. G. Balev,$^{a,b}$ Nelson Studart,$^{a}$ and P. Vasilopoulos$^{c}$

\smallskip

$^{a}${\small Departmento de Fisica, Universidade Federal de S\~{a}o Carlos,
13565-905 S\~{a}o Carlos, SP, Brazil}

$^{b}${\small Institute of Physics of Semiconductors, National Academy of
Sciences, 45 Pr. Nauky, Kiev 252650, Ukraine}

$^{c}${\small Concordia University, Department of Physics, 1455 de
Maisonneuve Blvd O, Montr\'{e}al, Qu\'{e}bec, Canada, H3G 1M8}\medskip
\end{center}

We present a {\it microscopic} treatment of edge magnetoplasmons (EMP's)
within the random-phase approximation for strong magnetic fields, low
temperatures, and filling factor $\nu =1(2)$, when a weak short-period
superlattice potential is imposed along the Hall bar. The modulation
potential modifies both the spatial structure and the dispersion relation of
the fundamental EMP and leads to the appearance of a novel gapless mode of
the fundamental EMP. For sufficiently weak modulation strengths the phase
velocity of this novel mode is almost the same as the group velocity of the
edge states but it should be quite smaller for stronger modulation. We
discuss in detail the spatial structure of the charge density of the
renormalized and the novel fundamental EMP's.

\smallskip

PACS\ \ 73.20.Dx, 73.20.Mf; 73.40.Hm

\section{INTRODUCTION}

The theory of edge magnetoplasmons (EMP's) plays a fundamental role in the
quantum Hall effect (QHE) due to a close interplay with time-resolved
transport experiments.\cite{exp} Classical\cite{classical} and quantum\cite
{quantum} models have been used to describe EMP modes through different wave
mechanisms at the edges of the two-dimensional electron system (2DES).
Recently a quasi-microscopic description for EMP's in the QHE regime,
embracing the edge-wave mechanisms mentioned above, has been proposed\cite
{balev97} that takes into account the lateral confining potential, the
structure of the Landau levels (LL) for integer values of the filling factor 
$\nu $, and the dissipation which conditions the propagation of the modes.
The lateral confining potential is flat in the interior of the channel and
smooth on the scale of the magnetic length $\ell _{0}$ but sufficiently
steep that the Landau level (LL) flattening can be neglected. The
theoretical framework was also extended to take into account nonlocal
contributions to the current density within the random-phase approximation.%
\cite{balev99}

There has been a great number of studies of magnetotransport properties of
2DES's modulated by one-dimensional (1D) lateral superlattices with large
period $a\agt 100$ nm.\cite{weiss89,cfsaw} The spectrum of magnetoplasmon
excitations has been studied in such systems as well \cite{cui}. Also, many
works have been devoted to magnetotransport and related phenomena in the
case of 2D lateral superlattices both in the regime of relatively weak
modulation \cite{thouless,macdonald} and for antidot arrays \cite{antidot},
where the 2DES cannot penetrate into the antidot region with higher
potential. Quite recently a superlattice field effect transistor was
designed in which the 2DES in a GaAs-based sample is subjected to a
atomically precise 1D potential with period of $15$ nm\cite{short} and
vicinal superlattices were produced with $a\approx 16$ nm.\cite{vicinal}
More recently, attention has been focused on transport of commensurate
composite fermions in weak periodic electrostatic potentials at the
half-filled LL.\cite{cfsaw}

In this work we employ the self-consistent field formalism, or random-phase
approximation (RPA), to study the effect of a 1D weak periodic modulation,
with period $a,$ on the fundamental EMPs in the quantum Hall system for LL
filling factors $\nu =1$ $(2)$ and low temperatures, $k_{B}T\ll \hbar
v_{g}/\ell _{0}$, where $v_{g}$ is the group velocity of edge states.
Motivated by recent results on the 2DES subjected to lateral superlattice
potentials with short period\cite{short,vicinal}, we restrict our study,
starting from Sec. III, to the short-period regime $a\ll 2\pi \ell_{0}$.
More precisely we will assume $\exp [-(\pi \ell _{0}/a)^{2}]\ll 1$.

In Sec. II we derive the integral equations for the wave charge density and
electrostatic potential at the edge of a periodically modulated channel. In
Sec. III, we present our result for the dispersion relations and spatial
structures of the fundamental EMPs. Finally, in Sec. IV we summarize our
major conclusions.

\section{Integral Equations for EMPs}

The non-interacting zero-thickness 2DEG, of width $W$ and length $L_{x}=L$,
in the presence of a strong magnetic field $B$ along the $z$ axis and under
a 1D periodic modulation, is described by the Hamiltonian $\hat{H}_{0}=\hat{h%
}^{0}+V_{s}(x),$ where $\hat{h}^{0}=[(\hat{p}_{x}+eBy/c)^{2}+\hat{p}%
_{y}^{2}]/2m^{*}+V_{y}$. The confining potential is flat in the interior of
the 2DES, ($V_{y}=0$) and parabolic at its edges, $V_{y}=m^{*}\Omega
^{2}(y-y_{r})^{2}/2$, $y\geq y_{r}$. We assume that $V_{y}$ is smooth on the
scale of $\ell _{0}$ such that $\Omega \ll \omega _{c}$, where $%
\omega_{c}=|e|B/m^{*}c$ is the cyclotron frequency. The 1D modulation $%
V_{s}(x)=V_{s}\cos (Gx)$ is a weak periodic potential with $G=2\pi /a$.
Within the RPA\ framework, the corresponding one-electron density matrix $%
\hat{\rho}$ obeys the equation of motion

\begin{equation}
i\hbar \frac{\partial \hat{\rho}}{\partial t}=[\hat{H}(t),\hat{\rho}]- \frac{%
i\hbar }{\tau }(\hat{\rho}-\hat{\rho}^{(0)}),  \label{1}
\end{equation}
where $\hat{H}(t)=\hat{H}_{0}+V(x,y,t)$ with

\begin{equation}
V(x,y,t)=e^{-i(\omega _{0}t-q_{x}x)}\sum_{l=-\infty }^{\infty }V_{l}(\omega
_{0},q_{x},y)e^{iGlx}+\text{c.c}.  \label{2}
\end{equation}
Without interaction the one-electron density matrix $\hat{\rho}^{(0)}$ is
diagonal, i.e., $<\alpha |\hat{\rho}^{(0)}|\beta >=f_{\alpha }\delta
_{\alpha \beta }$, where $f_{\alpha }=[1+exp((E_{\alpha
}-E_{F})/k_{B}T)]^{-1}$ is the Fermi-Dirac function; $\hat{H}_{0}|\alpha
>=E_{\alpha }|\alpha >$. Notice that $\tau \rightarrow \infty $ corresponds
to the collisionless case while a finite $\tau $ provides the possibility of
estimating roughly the influence of collisions. Both $\hat{H}_{0}$ and the
sum $\sum_{l=-\infty }^{\infty }V_{l}(\omega _{0},q_{x},y)e^{iGlx}$ are
periodic along $x$ with period $a$.

Equation (\ref{1}) can be solved by Laplace transforms.\cite{balev99} Taking
the trace of $\hat{\rho}$ with the electron density operator $e\delta ({\bf r%
}-\hat{{\bf r}})$ gives the wave charge density in the form

\begin{equation}
\rho (\omega _{0},x,y)=e^{iq_{x}x}\sum_{l=-\infty }^{\infty }
\rho_{l}(\omega _{0},q_{x},y)e^{iGlx}.  \label{3}
\end{equation}
The charge density $\rho_{l}(\omega ,q_{x},y)\exp [i(q_{x}+Gl)x]$ induces a
wave electric potential $\phi _{l}(\omega ,q_{x},y)\exp [i(q_{x}+Gl)x]$.
From Poisson's equation this is given as

\begin{equation}
\phi _{l}(\omega ,q_{x},y)=\frac{2}{\epsilon }\int_{-\infty }^{\infty
}dy^{\prime }K_{0}(|q_{x}+Gl||y-y^{\prime }|)\times \rho _{l}(\omega
,q_{x},y^{\prime }),  \label{4}
\end{equation}
where $\epsilon$ is the background dielectric constant, assumed spatially
homogeneous, and $K_{0}(x)$ is the modified Bessel function; $\phi $ and $%
\rho $ pertain to the 2D plane. Taking $|q_{x}|W\gg 1$, we can consider an
EMP along the right edge of the channel of the form $A(\omega
,q_{x},x,y)\exp [-i(\omega t-q_{x}x)]$ totally independent of the left edge,
where $A(\omega ,q_{x},x,y)$ is periodic along $x$ with period $a$.

In the absence of an external potential $V_{l}(\omega
,q_{x},y)=e\phi_{l}(\omega ,q_{x},y)$. As a result, considering large time
responses, $t \gg \tau $, after some straightforward calculations we obtain
the following integral equation for $\rho_{m}(\omega ,q_{x},y)$

\begin{eqnarray}
\rho _{m}(\omega ,q_{x},y) &=&\frac{2e^{2}}{\epsilon L}\sum_{\alpha ,\beta
}\sum_{l=-\infty }^{\infty }\int_{0}^{L}dx\ e^{-i(q_{x}+Gm)x}\psi _{\beta
}^{*}({\bf r})\psi _{\alpha }({\bf r})  \nonumber \\
&&\   \nonumber \\
&&\times \frac{f_{\beta }-f_{\alpha }}{E_{\beta }-E_{\alpha }+\hbar \omega
+i\hbar /\tau }\int d\tilde{{\bf r}}\ e^{i(q_{x}+Gl)\tilde{x}}\psi _{\alpha
}^{*}(\tilde{{\bf r}})\psi _{\beta }(\tilde{{\bf r}})  \nonumber \\
&&\   \nonumber \\
&&\times \int_{-\infty }^{\infty }dy^{\prime }\ K_{0}(|q_{x}+Gl||\tilde{y}%
-y^{\prime }|)\ \rho _{l}(\omega ,q_{x},y^{\prime }),  \label{4a}
\end{eqnarray}
where $\psi_{\alpha }=\langle {\bf r}\left| \alpha \right\rangle $ and we
dropped the subscript $0$ from $\omega_{0}$. For definiteness, we take $%
\omega >0$.

We consider low temperatures $T$ satisfying $\hbar v_{gn}\gg \ell _{0}k_{B}T$%
, where $v_{gn}$ is the group velocity of the edge states of $n$-th LL.
Furthermore we will assume the long-wavelength limit $q_{x}\ell _{0}\ll 1$,
which is well satisfied, e.g., for the fundamental EMP in the low-frequency
regime, $\omega \ll \omega _{c}$.\cite{balev97} Then, assuming that the
condition $Gv_{gn}\ll \omega _{c}$ is satisfied and comparing the terms
proportional to $f_{\beta ^{*}}$, for a given $n_{\beta ^{*}}$, of the
right-hand side (RHS) of Eq. (\ref{4a}), we conclude that the contribution
to the summation over $n_{\alpha }$ with $n_{\alpha }=n_{\beta ^{*}}$ is
much larger than any other term of this sum or the sum of all terms with $%
n_{\alpha }\neq n_{\beta ^{*}}$. The small parameter is $|\omega
-q_{x}v_{gn_{\beta ^{*}}}(k_{x\beta })|/\omega _{c}\ll 1$, where $%
v_{gn_{\beta ^{*}}}(k_{x\beta })$ is the group velocity of an occupied state 
$\left| n_{\beta ^{*}},k_{x\beta }\right\rangle $ of the $n_{\beta ^{*}}$
LL. The inequality above also implies that $q_{x}v_{g0}/\omega _{c}\ll 1$,
since $v_{g0}$ has typically the largest value among $v_{gn}$. Similar
results follow from an analysis of the terms proportional to $f_{\alpha
^{*}} $ in the summation over $n_{\beta }$ on the RHS of Eq. (\ref{4a}).
Hence, for $\omega \ll \omega _{c}$, $q_{x}v_{g0}\ll \omega _{c}$ and $%
Gv_{g0}\ll \omega _{c}$, the terms with $n_{\alpha }\neq n_{\beta }$ can be
neglected. Then the integral equation for the electron charge density $\rho
_{m}(\omega ,q_{x},y)$ becomes

\begin{eqnarray}
\rho _{m}(\omega ,q_{x},y) &=&\frac{e^{2}}{L}\sum_{n_{\alpha }=0}^{\bar{n}%
}\sum_{k_{x\alpha }}\sum_{k_{x\beta }}\sum_{l=-\infty }^{\infty
}\int_{0}^{L}dx\ e^{-i(q_{x}+Gm)x}\psi _{n_{\alpha }k_{x\beta }}^{*}({\bf r}%
)\psi _{n_{\alpha }k_{x\alpha }}({\bf r})  \nonumber \\
&&\times \frac{f_{n_{\alpha }k_{x\beta }}-f_{n_{\alpha }k_{x\alpha }}}{%
E_{n_{\alpha }k_{x\beta }}-E_{n_{\alpha }k_{x\alpha }}+\hbar \omega +i\hbar
/\tau }  \nonumber \\
&&\times \int d\tilde{{\bf r}}\ e^{i(q_{x}+Gl)\tilde{x}}\psi _{n_{\alpha
}k_{x\alpha }}^{*}(\tilde{{\bf r}})\psi _{n_{\alpha }k_{x\beta }}(\tilde{%
{\bf r}})\phi _{l}(\omega ,q_{x},\tilde{y}),  \label{5}
\end{eqnarray}
where $\bar{n}$ denotes the highest occupied Landau level (LL). For even $%
\nu $, the (RHS) of Eq. (\ref{5}) should be multiplied by 2, the spin
degeneracy factor; for $\nu $ even the spin-splitting is neglected. Equation
(\ref{5}), for $m=0,\pm 1,\pm 2,...$, gives a system of integral equations,
whose solution determines $\rho_{m}(\omega ,q_{x},y)$ in the RHS of Eq. (\ref
{3}).

Because $V_{s}(x)$ is assumed weak, the eigenfunctions $\psi _{n_{\alpha
}k_{x\alpha }}=\langle {\bf r}\left| n_{\alpha },k_{x\alpha }\right\rangle$
and the eigenvalues $E_{n_{\alpha }k_{x\alpha }}$ of $\hat{H}_{0}$ can be
evaluated by second-order perturbation theory. A straightforward calculation
leads to\cite{landau}

\begin{eqnarray}
\left| n_{\alpha },k_{x\alpha }\right\rangle &=&\left[ 1-\tilde{V}%
_{sn_{\alpha }}^{2}(k_{x\alpha })\right] \left| n_{\alpha },k_{x\alpha
}\right\rangle ^{\left( 0\right) }+\tilde{V}_{sn_{\alpha }}(k_{x\alpha
})\left[ \left| n_{\alpha },k_{x\alpha }-G\right\rangle ^{\left( 0\right)
}-\left| n_{\alpha },k_{x\alpha }+G\right\rangle ^{\left( 0\right) }\right] 
\nonumber \\
&&  \nonumber \\
&&+\tilde{V}_{sn_{\alpha }}^{2}(k_{x\alpha })\left[ \left| n_{\alpha
},k_{x\alpha }+2G\right\rangle ^{\left( 0\right) }+\left| n_{\alpha
},k_{x\alpha }-2G\right\rangle ^{\left( 0\right) }\right] .  \label{6}
\end{eqnarray}
Here we have introduced a dimensionless parameter $\tilde{V}_{sn}(k_{x})$
characterizing the strength of the periodic potential for the $n$-th LL,
near its edge, and given by

\begin{equation}
\tilde{V}_{sn}(k_{x})=\frac{V_{s}}{2\hbar Gv_{gn}(k_{x})} e^{-(G%
\ell_{0}/2)^{2}}L_{n}((G\ell _{0})^{2}/2),  \label{7}
\end{equation}
where $v_{gn}(k_{x})=\hbar ^{-1}\partial E_{n}(k_{x})/\partial k_{x}$ is the
group velocity of a state in the edge region of the $n$-th LL and $L_{n}(x)$
is the Laguerre polynomial. Due to the smoothness of the confining potential
on the $\ell _{0}$ scale, the unperturbed eigenfunctions are well
approximated by $\psi _{n_{\alpha }k_{x\alpha }}^{(0)}\equiv \langle {\bf r}%
\left| n_{\alpha },k_{x\alpha }\right\rangle ^{\left( 0\right) }\equiv \psi
_{\alpha }^{(0)}({\bf r})\approx e^{ik_{x}x}\Psi _{n}(y-y_{0})/\sqrt{L}$,
where $\Psi _{n}(y)$ is the harmonic oscillator function. Because we have
used the condition $\omega _{c}\gg Gv_{gn}$ to obtain Eq. (\ref{6}), the
small ``nonresonance'' contributions with $n_{\beta }\neq n_{\alpha }$ can
be neglected. In the edge region the evaluation of the eigenvalue $%
E_{n_{\alpha }k_{x\alpha }}\equiv E_{n_{\alpha }}(k_{x\alpha })$, by
perturbation theory, shows that the first-order correction $%
E_{n_{\alpha}}^{(1)}(k_{x\alpha })$ vanishes. As for the second-order
correction $E_{n_{\alpha}}^{(2)}(k_{x\alpha })$, the main ``resonance''
contributions to it, with $n_{\beta}=n_{\alpha }$, are mutually cancelled
due to imposed conditions. Then it can be shown that $E_{n_{\alpha
}}(k_{x\alpha })$ in the edge region can be well approximated by the
zero-order term, i.e., $E_{n_{\alpha }}(k_{x\alpha })\approx E_{n_{\alpha
}}^{(0)}(k_{x\alpha })$. The energy spectrum of the $n $-th LL, $E_{\alpha
}^{(0)}\approx (n+1/2)\hbar \omega _{c}+m^{*}\Omega ^{2}(y_{0}-y_{r})^{2}/2,$
leads to the group velocity of the edge states $v_{gn}=\partial
E_{n}(k_{r}+k_{e}^{(n)})/\hbar \partial k_{x}=\hbar \Omega
^{2}k_{e}^{(n)}/m^{*}\omega _{c}^{2}$ with characteristic wave vector $%
k_{e}^{(n)}=(\omega _{c}/\hbar \Omega )\sqrt{2m^{*}\Delta _{Fn}}$, $\Delta
_{Fn}=E_{F}-(n+1/2)\hbar \omega _{c}$, where $E_{F}$ is the Fermi energy.
The edge of the $n$-th LL is denoted by $y_{rn}=y_{r}+\ell
_{0}^{2}k_{e}^{(n)}=\ell _{0}^{2}k_{rn}$, where $k_{rn}=k_{r}+k_{e}^{(n)}$,
and $W=2y_{r0}$. We can also write $v_{gn}=cE_{en}/B$, where $E_{en}=\Omega 
\sqrt{2m^{*}\Delta _{Fn}}/|e|$ is the electric field associated with the
confining potential $V_{y}$ at $y_{rn}$. We have also introduced the wave
vector $k_{r}=y_{r}/\ell _{0}^{2}$. The typical width of the edge region for
the $n$-th LL can be estimated here as $\eta \ell _{0}^{2}k_{e}^{(n)}\gg
\ell _{0}$, where $\eta \ll 1$. For all occupied LLs, we assume that $%
k_{e}^{(n)}\gg G\gg 1/\ell _{0}$. Since in Eqs. (\ref{4a}) and (\ref{5}) the
significant eigenstates are localized along the $y$ direction near the right
edge of the channel, i.e., with $y_{0}(k_{x})>y_{r}$, we have considered
only these eigenstates in Eqs. (\ref{6}) and (\ref{7}). Moreover, it follows
from Eq. (\ref{5}) that the main contributions come from $k_{x}\approx
k_{rn} $ and $k_{x}\approx (k_{rn}\pm G)$; thus for the applicability of the
perturbation theory here it is sufficient to assume that $\tilde{V}%
_{sn}(k_{rn})\equiv \tilde{V}_{sn}\ll 1$.

\section{Fundamental EMPs for $\nu =1(2)$}

We first consider the case $\nu =1$ and then indicate how the results change
for $\nu =2$. For $\nu =1,$ we have $\bar{n}=0$ in Eq. (\ref{5}). We will
look for gapless edge modes, with $\omega \rightarrow 0$ for $%
q_{x}\rightarrow 0$, and assume that $1\gg \tilde{V}_{s0}\agt \exp [-(G\ell
_{0}/2)^{2}]$, where $\tilde{V}_{s0}=(V_{s}/2\hbar Gv_{g0})\exp [-(G\ell
_{0}/2)^{2}]$, which implies the short-period regime, $G\ell _{0}\gg 1$.
>From Eq. (\ref{5}) for $m=0,$ we can write the integral equation for $\rho
_{0}(\omega ,q_{x},y)$ in the form

\begin{equation}
\rho _{0}(\omega ,q_{x},y)=\left[ \hat{F}_{1}+\hat{F}_{2}\right]
\int_{-\infty }^{\infty }dy^{\prime }\ K_{0}(|q_{x}||\tilde{y}-y^{\prime
}|)\ \rho _{0}(\omega ,q_{x},y^{\prime }),  \label{8}
\end{equation}
where the integral functional $\hat{F}_{1}$ is given as

\begin{eqnarray}
\hat{F}_{1} &=&\frac{e^{2}}{\pi \hbar \epsilon }\int_{-\infty }^{\infty
}dk_{x\alpha }\Pi (y,k_{x\alpha })\frac{f_{0,k_{x\alpha
}-q_{x}}-f_{0,k_{x\alpha }}}{\omega -v_{g0}(k_{x\alpha })q_{x}+i/\tau } 
\nonumber \\
&&\   \nonumber \\
&&\times \int_{-\infty }^{\infty }d\tilde{y}\ \{\Pi (\tilde{y},k_{x\alpha })+%
\tilde{V}_{s0}^{2} [\Pi (\tilde{y},k_{x\alpha }-G)+\Pi (\tilde{y}%
,k_{x\alpha}+G)]\},  \label{9}
\end{eqnarray}
with $\Pi (y,k_{x\alpha })=\left| \Psi _{0}(y-y_{0}(k_{x\alpha
}))\right|^{2} $. The integral functional $\hat{F}_{2}$ is given as

\begin{eqnarray}
\hat{F}_{2} &=&\frac{e^{2}}{\pi \hbar \epsilon }\int_{-\infty }^{\infty
}dk_{x\alpha }\tilde{V}_{s0}^{2}[\Pi (y,k_{x\alpha }-G)+ \Pi
(y,k_{x\alpha}+G)]  \nonumber \\
&& \   \nonumber \\
&&\times \frac{f_{0,k_{x\alpha }-q_{x}}-f_{0,k_{x\alpha }}}{\omega
-v_{g0}(k_{x\alpha })q_{x}+i/\tau }\int_{-\infty }^{\infty } d\tilde{y}\ \Pi
(\tilde{y},k_{x\alpha }).  \label{10}
\end{eqnarray}
In Eqs. (\ref{8})-(\ref{10}), for $q_{x}\rightarrow 0$, we can make
approximation $(f_{0,k_{x\alpha }-q_{x}}-f_{0,k_{x\alpha }})\approx
q_{x}\delta (k_{x\alpha}-k_{r0})$. After integration over $k_{x\alpha },$ we
obtain

\begin{eqnarray}
\rho _{0}(\omega ,q_{x},\bar{y}) &=&\frac{e^{2}}{\pi \hbar \epsilon }\frac{%
q_{x}}{\tilde{\omega}}{\Large \{}\Psi _{0}^{2}(\bar{y})\ \int_{-\infty
}^{\infty }d\bar{y}^{\prime }\text{ }\left[ \Psi _{0}^{2}(\bar{y}^{\prime })+%
\tilde{V}_{s0}^{2}[\Psi _{0}^{2}(\bar{y}^{\prime }+G\ell _{0}^{2})+\Psi
_{0}^{2}(\bar{y}^{\prime }-G\ell _{0}^{2})]\right]  \nonumber \\
&&+\ \tilde{V}_{s0}^{2}\left[ \Psi _{0}^{2}(\bar{y}+G\ell _{0}^{2})+\Psi
_{0}^{2}(\bar{y}-G\ell _{0}^{2})\right] \int_{-\infty }^{\infty }d\bar{y}%
^{\prime }\Psi _{0}^{2}(\bar{y}^{\prime }){\Large \}}  \nonumber \\
&&\times \int_{-\infty }^{\infty }d\bar{y}^{\prime \prime }\text{ }%
K_{0}(|q_{x}||\bar{y}^{\prime }-\bar{y}^{\prime \prime }|)\ \rho _{0}(\omega
,q_{x},\bar{y}^{\prime \prime }),  \label{11}
\end{eqnarray}
where $\bar{y}=y-y_{r0}$, and $\tilde{\omega}=\omega -q_{x}v_{g0}+i/\tau $.
In order to simplify the notation we take $\rho _{i}(\omega ,q_{x},y)\equiv
\rho _{i}(\omega ,q_{x},\bar{y})$, $i=0,\pm 1,..$.

Similarly, omitting minor terms in Eq. (\ref{5}) we obtain, for $m=1$,

\begin{eqnarray}
\rho _{1}(\omega ,q_{x},\bar{y}) &=&\frac{e^{2}}{\pi \hbar \epsilon } \tilde{%
V}_{s0}\frac{q_{x}}{\tilde{\omega}}\ \Psi _{0}(\bar{y})[\Psi _{0}(\bar{y}%
+G\ell _{0}^{2})- \Psi _{0}(\bar{y}-G\ell _{0}^{2})]\   \nonumber \\
&&\   \nonumber \\
&&\times \int_{-\infty }^{\infty }d\tilde{y}\ \Psi _{0}^{2}(\tilde{y}%
)\int_{-\infty }^{\infty }d\bar{y}^{\prime }\ K_{0}(|q_{x}||\tilde{y}-\bar{y}%
^{\prime }|)\ \rho _{0}(\omega ,q_{x},\bar{y}^{\prime }).  \label{12}
\end{eqnarray}
From Eq. (\ref{5}) we find, for $m=-1$, $\rho_{-1}(\omega ,q_{x},y) \equiv
\rho_{1}(\omega,q_{x},y)$.

The general solution of the linear homogeneous integral equation, Eq. (\ref
{11}), can be sought in the form

\begin{equation}
\rho _{0}(\omega ,q_{x},y)=\rho _{0}^{(0)}(\omega ,q_{x})\Psi _{0}^{2}(\bar{y%
})+\rho _{0}^{(1)}(\omega ,q_{x})[\Psi _{0}^{2}(\bar{y}+G\ell _{0}^{2})+\Psi
_{0}^{2}(\bar{y}-G\ell _{0}^{2})].  \label{13}
\end{equation}
Substituting Eq. (\ref{13}) into Eq. (\ref{11}) and equating the
coefficients of $\Psi _{0}^{2}(\bar{y})$ and $[\Psi _{0}^{2}(\bar{y}+G\ell
_{0}^{2})+\Psi _{0}^{2}(\bar{y}-G\ell _{0}^{2})]$ on both sides of Eq. (\ref
{11}), we obtain two linear homogeneous equations for $\rho_{0}^{(i)}(\omega
,q_{x})$, $i=0,1$:

\begin{eqnarray}
\rho _{0}^{(0)} &=&\frac{e^{2}}{\pi \hbar \epsilon } \frac{q_{x}}{\tilde{%
\omega}}\{[a_{00}(q_{x})+ 2\tilde{V}_{s0}^{2}\
a_{00}^{00}(q_{x},G\ell_{0}^{2})] \rho _{0}^{(0)}  \nonumber \\
&& \   \nonumber \\
&&+2[a_{00}^{00}(q_{x},G\ell _{0}^{2})+\tilde{V}_{s0}^{2}\
(a_{00}(q_{x})+a_{00}^{00}(q_{x},2G\ell _{0}^{2}))]\rho _{0}^{(1)}\}
\label{14}
\end{eqnarray}
and

\begin{equation}
\rho _{0}^{(1)}=\frac{e^{2}}{\pi \hbar \epsilon }\tilde{V}_{s0}^{2}\frac{%
q_{x}}{\tilde{\omega}}\{a_{00}(q_{x})\rho
_{0}^{(0)}+2a_{00}^{00}(q_{x},G\ell _{0}^{2})\rho _{0}^{(1)}\},  \label{15}
\end{equation}
where the coefficients $a_{nn}^{mm}$ are given by\cite{balev99}

\begin{equation}
a_{nn}^{mm}(q_{x},\Delta y)=\int_{-\infty }^{\infty }\int_{-\infty }^{\infty
}dx\ dx^{\prime }\Psi _{n}^{2}(x)\Psi _{m}^{2}(x^{^{\prime
}})K_{0}(|q_{x}||x-x^{\prime }+\Delta y|).  \label{15a}
\end{equation}
Here $a_{nn}^{mm}(q_{x},\Delta y)=a_{mm}^{nn}(q_{x},\Delta y)$, $%
a_{nn}^{mm}(q_{x},\Delta y)=a_{nn}^{mm}(q_{x},-\Delta y)$ and $%
a_{00}^{00}(q_{x},0)=a_{00}(q_{x})$. We will assume $2\pi q_{x}\ell
_{0}^{2}/a\ll 1$. Notice that $a_{00}(q_{x})\approx \ln (1/q_{x}\ell
_{0})+3/4$ and, for $\Delta y/\ell _{0}\gg 1$, $a_{00}^{00}(q_{x},\Delta
y)\approx \ln (2/q_{x}\ell _{0})-\gamma -\ln (\Delta y/\ell _{0})\approx \ln
(1/q_{x}\Delta y)+0.1$, where $\gamma $ is the Euler constant.

\subsection{Dispersion Relations}

The dimensionless frequencies $\omega^{\prime }=\tilde{\omega}%
/(e^{2}q_{x}/\pi \hbar \epsilon )$ of the branches resulting from the
determinantal solution of the two linear homogeneous equations for $%
\rho_{0}^{(i)}(\omega ,q_{x}),$ Eqs. (\ref{14}) and (\ref{15}), are given by

\begin{eqnarray}
\omega _{\pm }^{\prime } &=&{\frac{1}{2}}a_{00}(q_{x})+2\tilde{V}_{s0}^{2}\
a_{00}^{00}(q_{x},G\ell _{0}^{2}) \pm {\frac{1}{2}}a_{00}(q_{x})  \nonumber
\\
&&\times \{1+8\tilde{V}_{s0}^{2}\ a_{00}^{-1}(q_{x})[a_{00}^{00}(q_{x},G\ell
_{0}^{2})+\tilde{V}_{s0}^{2}\
(a_{00}(q_{x})+a_{00}^{00}(q_{x},2G\ell_{0}^{2}))]\}^{1/2}\;.  \label{16}
\end{eqnarray}

From Eq. (\ref{16}), it follows that the effect of the modulation potential
on the fundamental EMP is quite strong. Apart from the renormalization of
the fundamental EMP of $n=0$ LL with dispersion

\begin{equation}
\omega _{+}^{\prime }\approx a_{00}(q_{x})+4\tilde{V}_{s0}^{2}\
a_{00}^{00}(q_{x},G\ell _{0}^{2})  \label{17}
\end{equation}
it leads to the existence of a {\it novel} fundamental EMP of $n=0$ LL with
dispersion

\begin{equation}
\omega _{-}^{\prime }\approx 2\tilde{V}_{s0}^{4}\ a_{00}^{-1}(q_{x})\ \{2\
[a_{00}^{00}(q_{x},G\ell _{0}^{2})]^{2}-a_{00}(q_{x})\
[a_{00}(q_{x})+a_{00}^{00}(q_{x},2G\ell _{0}^{2})]\}.  \label{18}
\end{equation}
Substituting the coefficients into Eq.(\ref{18}), we obtain the dispersion
relation (DR) of the novel fundamental EMP

\begin{equation}
\omega _{-}\approx [v_{g0}-\frac{12}{\epsilon }\tilde{V}_{s0}^{4}\sigma
_{yx}^{0}\ln (G\ell _{0})]q_{x}-i/\tau ,  \label{19}
\end{equation}
where $\sigma _{yx}^{0}=\nu e^{2}/2\pi \hbar $ and $\nu =1(2).$ Note that
for $\tilde{V}_{s0}\leq 10^{-1}$ and $v_{g0}\geq 10^{6}$ cm/s, which is a
typical value in GaAs-based heterostructures, the correction in the phase
velocity of this novel mode should be quite small. However, in general, e.g.
for a slightly larger $\tilde{V}_{s0}$ and a slightly smaller $v_{g0}$, this
contribution should be taken into account and can lead to a substantial
decrease of the phase velocity of the novel fundamental EMP from its maximum
possible value $v_{g0}$. We point out that this contribution to Re $%
\omega_{-}$ stems from the electron-electron interaction and the strength of
the periodic modulation as well. Eq. (\ref{19}) is valid for

\begin{equation}
\tilde{V}_{s0}^{2}\gg \frac{1}{3\ln (G\ell _{0})}\frac{\epsilon v_{g0}}{%
2\sigma _{yx}^{0}}\left( \frac{2\pi }{k_{e}^{(0)}a}\right) ^{2}.  \label{19a}
\end{equation}
Notice that the RHS of the inequality (\ref{19a}) is typically very small.
Under this condition, the second order correction in the group velocity $%
v_{g0}(k_{r0})=v_{g0}[1-2\tilde{V}_{s0}^{2}(2\pi /k_{e}^{(0)}a)^{2}]$ can be
neglected. As discussed above this is equivalent to neglecting $%
E_{0}^{(2)}(k_{x\alpha })$.

From Eq. (\ref{17}), the DR for the renormalized fundamental EMP can be
written as

\begin{equation}
\omega_{+}\approx q_{x}v_{g0}+\frac{2}{\epsilon }\sigma _{yx}^{0}q_{x}
\{\ln(1/q_{x}\ell _{0})+\frac{3}{4}+4\tilde{V}_{s0}^{2}\ \ln
(1/(q_{x}G\ell_{0}^{2}))\}-i/\tau .  \label{20}
\end{equation}
The term $\propto \tilde{V}_{s0}^{2}$ shows a strong renormalization of the
fundamental EMP that depends on the strength and the period of the
modulation, for given value of $q_{x}$.

For a GaAs-based 2DEG and negligible dissipation, the dispersion laws for
the renormalized, by the superlattice potential and intra-LL Coulomb
coupling, fundamental EMP and for the novel fundamental EMP, caused by the
periodic modulation $V_{s}(x)$, are shown in Fig. 1 by the top and bottom
solid curves, respectively. The DRs corresponding to the $\omega _{-}$ and $%
\omega _{+}$ modes here are obtained using Eqs. (\ref{19}) and (\ref{20}),
respectively. For the assumed parameters these equations very well
approximate the exact DRs given by Eq. (\ref{16}). For the sake of
comparison, the dashed curve in Fig. 1 shows the fundamental EMP of $n=0$ LL
in the absence of superlattice potential.

As we have discussed, the renormalization effect involves essentially
intra-LL Coulomb coupling. In Fig. 1 the parameters are $m^{*}\approx
6.1\times 10^{-29}$ g, $\epsilon \approx 12.5$, and $\Omega \approx
7.8\times 10^{11}$ s$^{-1}$.\cite{muller92} Assuming $\nu =1$ and $B=9$ T,
these parameters lead to $\omega _{c}/\Omega \approx 30$. Here $\omega
_{*}=e^{2}/\pi \hbar \epsilon \ell _{0}\approx 0.3\omega _{c}\approx 7\times
10^{12}$ s$^{-1}$ is a characteristic frequency. We have also assumed $%
\Delta _{F0}=\hbar \omega _{c}/2$, $\tilde{V}_{s0}=\exp (-2)\ll 1$, and $%
a=\pi \ell _{0}/\sqrt{2}$. This gives $v_{g0}\approx \Omega \ell _{0}\approx
6.5\times 10^{5}\ $cm/s, $a\approx 18.5$ nm, $V_{s}=2.9$ meV. Observe that
for these parameters, the second term in the RHS of Eq. (\ref{19}) is more
than 50 times smaller than the first term, $v_{g0}$. Hence, the curve $%
\omega =q_{x}v_{g0}$ will practically coincide with the solid curve at the
bottom of Fig. 1.

The dispersion laws corresponding to the $\omega_{+}$ and $\omega_{-}$
modes, given by Eq. (\ref{16}), are depicted in Fig. 2 by the top and bottom
curves respectively. The solid, dot-dashed, and dashed curves correspond to $%
\tilde{V}_{s0}=0.3,\;0.2$, and $0.1$, respectively. The data of the bottom
curves were multiplied by the factor 30. The parameters are the same as in
Fig. 1 except $\Delta_{F0}=\hbar \omega _{c}/8$, $a\approx 20.6$ nm. This
leads to $v_{g0}\approx 3.25\times 10^{5}\ $cm/s and to modulation strengths 
$V_{s}\approx 2.18$ meV, $1.09$ meV, and $0.73$ meV for $\tilde{V}%
_{s0}=0.3,\;0.2$, and $0.1$, respectively. Notice that in this case $\exp[%
-(G\ell _{0}/2)^{2}]\approx 0.2$. It is seen that by varying the amplitude $%
V_{s}$ of the periodic potential, strong modifications in the DR of the
fundamental modes can occur. We observe that the phase velocity of the novel
EMP decreases from its maximum value $v_{g0}$ by increasing $\tilde{V}_{s0}$%
. It can be shown that the DRs given by Eqs. (\ref{19}) and (\ref{20}) still
represent well all curves in Fig. 2.

\subsection{Spatial Structure}

If we substitute Eq. (\ref{17}) into Eq. (\ref{15}), we obtain $\rho
_{0}^{(1)}(\omega _{+},q_{x})/\rho _{0}^{(0)}(\omega _{+},q_{x})\approx 
\tilde{V}_{s0}^{2}$. As a consequence, only a small distortion of the edge
charge occurs at $\bar{y}=\pm G\ell _{0}^{2}$, in comparison with the usual
one at $\bar{y}=0$. Furthermore substituting Eq. (\ref{18}) in Eq. (\ref{15}%
), we obtain $\rho _{0}^{(0)}(\omega _{-},q_{x})/\rho
_{0}^{(1)}(\omega_{-},q_{x}) \approx -2[\ln (1/q_{x}\ell _{0})-\ln (G\ell
_{0})]/[\ln (1/q_{x}\ell _{0})+3/4]>-2.$ This means that the amplitude of
the edge charges localized at $\bar{y}=\pm G\ell _{0}^{2}$ has absolute
value approximately twice smaller in comparison with that of the charge
distortion localized at $\bar{y}=0$; in addition, it has the opposite sign.
Furthermore, the ratio of amplitudes of the novel fundamental EMP is
independent on $V_{s0}$ for the assumed conditions, while in the case of the
renormalized fundamental EMP such ratio tends to zero. The same results hold
for $\nu =2$. Thus the novel mode has a spatial structure quite different
both from the spatial structure of the fundamental EMP (i.e., in the absence
of modulation) and from the renormalized mode.

We proceed now to evaluate the charge density $\rho _{1}(\omega ,q_{x},y)$
induced by $\rho _{0}(\omega ,q_{x},y)$ for the two new branches: the
renormalized fundamental EMP and the novel fundamental EMP. For both
fundamental EMPs we obtain, from Eq. (\ref{12}),

\begin{equation}
\rho _{1}(\omega ,q_{x},y)=\rho _{1}(\omega ,q_{x})\Psi _{0}(\bar{y})[\Psi
_{0}(\bar{y}+G\ell _{0}^{2})-\Psi _{0}(\bar{y}-G\ell _{0}^{2})],  \label{37}
\end{equation}
where, using Eq. (\ref{15}), we find

\begin{equation}
\rho _{1}(\omega ,q_{x})=\rho _{0}^{(1)}(\omega ,q_{x})/\tilde{V}_{s0}.
\label{38}
\end{equation}
Then for the renormalized fundamental EMP the relative amplitude $\xi
_{+}\equiv (\omega _{+},q_{x},y)/\rho _{0}(\omega _{+},q_{x},y)\approx 
\tilde{V}_{s0}\ \exp [-(G\ell _{0}/2)^{2}]$, where the small factor $\exp
[-(G\ell _{0}/2)^{2}]\alt \tilde{V}_{s0}$ comes from the exponentially small
overlapping of the wave functions in the products of Eq. (\ref{37}).
Similarly, for the novel fundamental EMP, $\omega_{-}$, we obtain that

\begin{equation}
\frac{\rho _{1}(\omega _{-},q_{x})}{\rho _{0}^{(0)}(\omega _{-},q_{x})}%
\approx -\frac{a_{00}(q_{x})}{2\tilde{V}_{s0}\ a_{00}^{00}(q_{x},G\ell
_{0}^{2})}\rightarrow -\frac{1}{2\tilde{V}_{s0}},  \label{39}
\end{equation}
where the limit holds for $q_{x}\rightarrow 0$. Now the relative amplitude
is $\xi _{-}\equiv \rho _{1}(\omega _{-},q_{x},y)/\rho _{0}(\omega
_{-},q_{x},y)=\exp [-(G\ell _{0}/2)^{2}]/2\tilde{V}_{s0}$. Hence, $\xi _{-}$
lies in the interval $[0.1,$ $1]$, i.e., the amplitude of oscillations of
the charge distortion $\propto \rho_{\pm 1}(\omega ,q_{x},y)$ can be of the
same order of magnitude as that $\propto \rho_{0}(\omega ,q_{x},y)$. A
further treatment of Eq. (\ref{5}) for $m=2$ (and $\bar{n}=0$) shows that
the charge distortions $\rho_{2}(\omega ,q_{x},y)=\rho_{-2}(\omega ,q_{x},y)$
as compared with $\rho_{0}(\omega ,q_{x},y)$ for these two new branches have
an additional small factor $\propto \tilde{V}_{s0}\exp [-3(G\ell _{0}/2)^{2}]%
\alt
\tilde{V}_{s0}^{4}$ with respect to the relative strength $\xi _{\pm }$ of $%
\rho _{\pm 1}(\omega ,q_{x},y)$. Therefore, terms with $|l|\geq 2$ in Eq. (%
\ref{3}) can be neglected both for the renormalized fundamental EMP and the
novel fundamental EMP. As a result, from Eq. (\ref{3}), we obtain
straightforwardly the dimensionless form factors, $\rho _{\pm }(x,y)\equiv
\rho _{\pm }=\sqrt{\pi }\ell _{0}\ \exp (-iq_{x}x)\ \rho (\omega _{\pm
},x,y)/\rho _{0}^{(0)}(\omega _{\pm },q_{x})$, as

\begin{eqnarray}
\rho_{+}(x,y) &=&\sqrt{\pi }\ell _{0}\{\Psi _{0}^{2}(\bar{y})+ \tilde{V}%
_{s0}^{2}\ [\Psi _{0}^{2}(\bar{y}+G\ell _{0}^{2})+ \Psi_{0}^{2}(\bar{y}%
-G\ell _{0}^{2})]  \nonumber \\
&& \   \nonumber \\
&&+2\tilde{V}_{s0}\ \cos (Gx)\Psi _{0}(\bar{y}) [\Psi _{0}(\bar{y}%
+G\ell_{0}^{2})-\Psi _{0}(\bar{y}-G\ell _{0}^{2})]\}  \label{21}
\end{eqnarray}
for the renormalized fundamental mode, and

\begin{eqnarray}
\rho _{-}(x,y) &=&\sqrt{\pi }\ell _{0}\{\Psi _{0}^{2}(\bar{y})-\frac{1}{2}\
[\Psi _{0}^{2}(\bar{y}+G\ell _{0}^{2})+\Psi _{0}^{2}(\bar{y}-G\ell _{0}^{2})]
\nonumber \\
&&\   \nonumber \\
&&-\frac{1}{\tilde{V}_{s0}}\ \cos (Gx)\Psi _{0}(\bar{y})[\Psi _{0}(\bar{y}%
+G\ell _{0}^{2})-\Psi _{0}(\bar{y}-G\ell _{0}^{2})]\}  \label{22}
\end{eqnarray}
for the novel fundamental mode. In order to exhibit explicitly the $x$
dependence of the form factors, for the same parameters as used in Fig. 1,
we show them in Fig. 3 for $x_{0}^{(m)}=a(m+1/2)/2$, $m=0,\pm 1,\pm 2,...$,
with $\cos (2\pi x_{0}^{(m)}/a)=0$, and for $x_{1}^{(m)}=m\ a$, with $\cos
(2\pi x_{1}^{(m)}/a)=1$. The solid and dotted curves show $\rho _{+}(x,y)$
as a function of $Y=\bar{y}/\ell _{0}$ for $x_{1}^{(m)}=m\ a$ and $%
x_{0}^{(m)}=a(m+1/2)/2$, respectively. We see that the dotted curve is
exactly symmetrical with respect to $Y=0$ axis. The deviations of the solid
curve from this form come from contributions to the form factor that are
commensurate with the unidirectional modulation. Also in Fig. 3,\ $\rho
_{-}(x,y)$ is shown by dot-dashed and dashed curves for $x_{1}^{(m)}$ and $%
x_{0}^{(m)}$, respectively. Notice that the dashed curve is symmetric while
the dot-dashed curve is clearly asymmetric. In Fig. 4, we present results
for the charge densities of the renormalized and novel fundamental EMPs for
the same parameters that are used to obtain the solid curves in Fig. 2. That
is, in Fig. 4 we use $\tilde{V}_{s0}=0.3$ and $G\ell _{0}\approx 2.53$ in
Eqs. (\ref{21}) and (\ref{22}).

\section{Conclusions}

We have presented a fully {\it microscopic} model for EMPs in the RPA
framework valid for integer $\nu =1$ and $2$ in the case of an applied 1D
weak modulation $V_{s}(x)=V_{s}\cos (2\pi x/a)$, and confining potentials
that are smooth on the $\ell _{0}$ scale but still sufficiently steep at the
edges that LL flattening can be neglected.\cite{flattening} The model also
takes into account nonlocal responses and incorporates only very weak
dissipation. The main results of the present work are as follows.

i) The strength of the periodic modulation, if not too small, reshapes
noticeably the spatial structure of the usual fundamental EMP of $n=0$ LL 
\cite{balev99}, normal and parallel to the edge, and substantially modifies
the dispersion relation leading to a renormalized fundamental EMP of $n=0$
LL. For instance, in Fig. 1, we have seen that the group velocity of the
renormalized fundamental EMP is more than $4\%$ greater than that of the
fundamental EMP without modulation for $q_{x}\ell _{0}=0.8\times 10^{-2}$.
Therefore, in time-resolved experiments, the periodic potential will imply a
modulation of the propagation time of the signal due the renormalized
fundamental EMP. As we have seen, this renormalization depends on the
strength and period of the modulation potential.

ii) The strength of the periodic modulation, even quite small, leads to the
appearance of the {\it novel} fundamental EMP with acoustical dispersion
relation and phase velocity typically equal, in a GaAs-based sample, to the
group velocity of the edge states, $v_{g0}$, independent of $V_{s}$ and $a$,
if $\tilde{V}_{s0}=(V_{s}a/4\pi \hbar v_{g0})\exp [-(\pi \ell
_{0}/a)^{2}]\leq 10^{-1}$ and $v_{g0}\geq 10^{6}$ cm/s. That is, this holds
for a sufficiently weak periodic modulation. However, already for $\tilde{V}%
_{s0}\approx 0.2$ and $v_{g0}\alt 10^{6}$ cm/s, the phase velocity of the
novel fundamental EMP can be substantially smaller than $v_{g0}$, as one can
see in Fig. 2, due to the combined effect of a short-period lateral
superlattice and the electron-electron interaction. In addition, its spatial
structure is strongly dependent on both $\tilde{V}_{s0}$ and $a$. The
spatial structure is almost symmetric with respect to the edge of the $n=0$
LL for very weak periodic modulation. However, for not too weak $V_{s}(x)$,
it becomes substantially asymmetric for some regions of $x$, as one can see
by the dot-dashed curve in Figs. 3 and 4. We have also obtained that in the
latter case the contributions to the spatial structure of the novel
fundamental EMP that are commensurate with the periodic modulation can be of
the same order of magnitude as those that are independent of $x$.

iii) The measurement of the velocity of the novel EMP, due to its
independence of the modulation parameters, in a wide range of them, can be a
useful tool for obtaining directly the group velocity of edge states.
Furthermore, a qualitative analysis, using results of previous studies\cite
{balev97} as well the above findings, shows that the dominant contribution
to the damping rate of the novel EMP is absent. Then we may speculate that
the damping rate of the novel mode could be rather small.

\acknowledgements

This work was supported by the Brazilian FAPESP, grants no. 98/10192-2 and
no. 95/0789-3, and by Canadian NSERC grant no. OGP0121756. O. G. B. thanks
partial support by the Ukrainian SFFI grant no. 2.4/665 and N. S. is
grateful to CNPq for a research fellowship.

\bigskip

\begin{center}
FIGURE CAPTIONS
\end{center}

\medskip

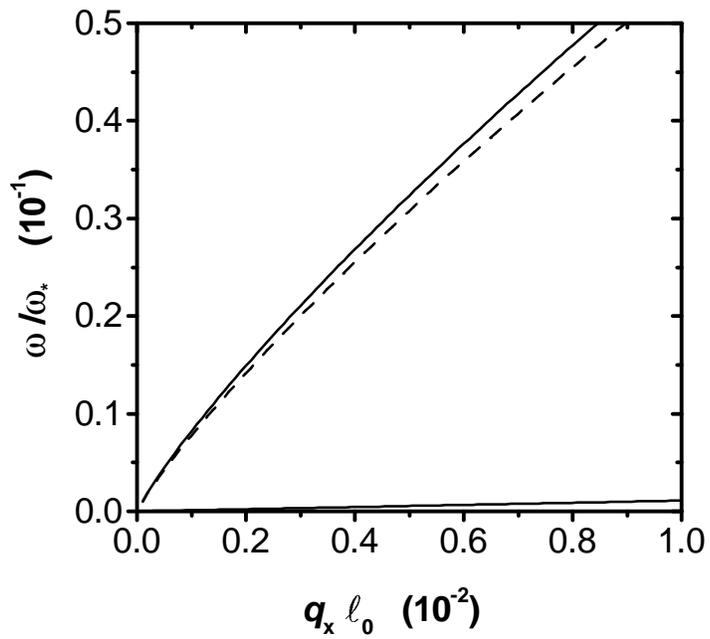
\noindent Fig. 1. Dispersion relations for $\omega _{\pm }$ modes, in units
of $\omega _{*}=e^{2}/\pi \hbar \epsilon \ell _{0}$, for $\nu =1$ and $B=9$
T. The upper solid curve is the renormalized fundamental EMP of the $n=0$ LL
and the lower solid one is the novel fundamental EMP, due to the effect of
the modulation potential. The dashed curve is the fundamental EMP in the
absence of the modulation. The parameters of a GaAs-based sample are given
in the text and we took $\Delta _{F0}=\hbar \omega _{c}/2$, $a\approx 18.5$
nm, and $\tilde{V}_{s0}=\exp (-2)$.

\medskip

\noindent Fig. 2. Dispersion relation of $\omega _{\pm }$ modes, in units of 
$\omega _{*}=e^{2}/\pi \hbar \epsilon \ell _{0}$, for values of the
modulation strength $\tilde{V}_{s0}=2.18$ meV$\;$(solid curve), $1.09$ meV
(dot-dashed) and $0.73$ meV (dashed) and $\nu =1$ and $B=9$ T. Top (bottom)
curves represent the dispersion laws for $\omega _{+}$ and $\omega _{-}$
modes, respectively. The values of $\omega _{-}$ are multiplied by $30$. The
parameters are the same as in Fig. 1, except $\Delta _{F0}=\hbar \omega
_{c}/8$, $a\approx 20.6$ nm. It follows that $v_{g0}\approx 3.25\times
10^{5}\ $cm/s and $\exp [-(G\ell _{0}/2)^{2}]\approx 0.2$.

\medskip

\noindent Fig. 3. Form factor for the fundamental EMP as a function of $Y=%
\bar{y}/\ell _{0}$, where $\bar{y}=y-y_{r0}$, and $y_{r0}$ is the edge of $%
n=0$ LL. The renormalized mode is indicated respectively by solid and dotted
curves for $x_{1}^{(m)}=m\ a$ and $x_{0}^{(m)}=a(m+1/2)/2$, with $m=0,\pm
1,\pm 2,...,$ and the novel EMP by dot-dashed and dashed curve for $%
x_{1}^{(m)}$ and $x_{0}^{(m)}$, respectively. The parameters used are the
same as in Fig. 1.

\medskip

\noindent Fig. 4. The same as in Fig. 3, but with the parameters used for
plotting the solid curves in Fig. 2.


\begin{references}
\bibitem{exp}  R. C. Ashoori, H. L. Stormer, L. N. Pfeiffer, K. W. Baldwin,
and K. West, Phys. Rev. B {\bf 45}, 3894 (1992); N. B. Zhitenev, R. J. Haug,
K. von Klitzing, and K. Eberl, Phys. Rev. Lett. {\bf 71}, 2292 (1993); {\it %
ibid}. Phys. Rev. B {\bf 49}, 7809 (1994); G. Ernst, R. J. Haug, J. Kuhl, K.
von Klitzing, and K. Eberl, Phys. Rev. Lett. {\bf 77}, 4245 (1996).

\bibitem{classical}  V. A. Volkov and S. A. Mikhailov, Zh. Eksp. Teor. Fiz. 
{\bf 94}, 217 (1988) [Sov. Phys. JETP {\bf 67}, 1639 (1988)]; I. L. Aleiner
and L. I. Glazman, Phys. Rev. Lett. {\bf 72}, 2935 (1994).

\bibitem{quantum}  X. G. Wen, Phys. Rev. B {\bf 43}, 11025 (1991); M. Stone,
Ann. Phys. (NY) {\bf 207}, 38 (1991); C. de Chamon and X. G. Wen, Phys. Rev.
B {\bf 49}, 8227 (1994); J. S. Giovanazzi, L. Pitaevskii, and S. Stringari,
Phys. Rev. Lett. {\bf 72}, 3230 (1994).

\bibitem{balev97}  O. G. Balev and P. Vasilopoulos, Phys. Rev. B {\bf 56},
13252 (1997); {\it ibid}. Phys. Rev. Lett. {\bf 81}, 1481 (1998); O. G.
Balev, P. Vasilopoulos, and Nelson Studart, J. Phys.: Condens. Matter {\bf 11%
}, 5143(1999).

\bibitem{balev99}  O. G. Balev and P. Vasilopoulos, Phys. Rev. B {\bf 59},
2807 (1999).

\bibitem{weiss89}  D. Weiss, K. von Klitzing, K. Ploog, and G. Weimann.
Europhys. Lett. {\bf 8}, 179 (1989); R. R. Gerhardts, D. Weiss, and K. von
Klitzing, Phys. Rev. Lett. {\bf 62}, 1173 (1989); R. W. Winkler, J. P.
Kotthaus, and K. Ploog, Phys. Rev. Lett. {\bf 62}, 1177 (1989).

\bibitem{cfsaw}  R. L. Willet, K. W. West, and L. N. Pfeiffer, Phys. Rev.
Lett. {\bf 78}, 4478 (1997); F. von Oppen, A. Stern, and B. I. Halperin,
Phys. Rev. Lett. {\bf 80}, 4494 (1998); S. D. M. Zwerschke and R. R.
Gerhardts, Phys. Rev. Lett. {\bf 83}, 2616 (1999); J. H. Smet, S. Jobst, K.
von Klitzing, D. Weiss, W. Wegscheider, and V. Umansky, Phys. Rev. Lett. 
{\bf 83}, 2620 (1999); R. L. Willet, K. W. West, and L. N. Pfeiffer, Phys.
Rev. Lett. {\bf 83}, 2624 (1999).

\bibitem{cui}  H. L. Cui, V. Fessatidis, and N. J. M. Horing, Phys. Rev.
Lett. {\bf 63}, 2598 (1989); S. M. Stewart and C. Zhang, Semicond. Sci.
Technol. {\bf 10}, 1541 (1995).

\bibitem{thouless}  D. J. Thouless, M. Kohmoto, M. P. Nightingale, and M.
den Nijs, Phys. Rev. Lett. {\bf 49}, 405 (1982); V. Gudmundsson, R. R.
Gerhardts, Phys. Rev. B {\bf 54}, 5223 (1996); T. Schlosser, K. Ensslin, J.
P. Kotthaus, and M. Holland, Semicond. Sci. Technol. {\bf 11}, 1582 (1996).

\bibitem{macdonald}  D. Pfannkuche, A. H. MacDonald, Phys. Rev. B {\bf 56},
7100 (1997); C. Albrecht, J. H. Smet, D. Weiss, K. von Klitzing, R. Henning,
Phys. Rev. Lett. {\bf 83}, 2234 (1999).

\bibitem{antidot}  D. Weiss, M. L. Roukes, A. Menschig, P. Grambow, K. von
Klitzing, and G. Weimann, Phys. Rev. Lett. {\bf 66}, 2790 (1991); S.
Ishizaka and T. Ando, Phys. Rev. B {\bf 56}, 15195 (1997); F. Nihey, K.
Nakamura, T. Takamasu, G. Kido, T. Sakon, and M. Motokawa, Phys. Rev. B {\bf %
59}, 14872 (1999).

\bibitem{short}  R. A. Deutschamann, A. Lorke, W. Wegscheider, M. Bichler,
and G. Abstreiter, {\it Proc. of EP2DS-13}, to appear in Physica E.

\bibitem{vicinal}  K. Tanaka, Y. Nakamura, and H. Sakaki, {\it Proc. of
EP2DS-13}, to appear in Physica E.

\bibitem{landau}  L. D. Landau and E. M. Lifshitz, {\it Quantum Mechanics},
(Pergamon Press, New York, 1975).

\bibitem{muller92}  G. Muller, D. Weiss, A. V. Khaetskii, K, von Klitzing,
S. Koch, H. Nickel, W. Schlapp, and R. Losch, Phys. Rev. B. {\bf 45}, 3932
(1992).

\bibitem{flattening}  D. B. Chklovskii, B. I. Shklovskii, and L. I. Glazman,
Phys. Rev. B {\bf 46}, 4026 (1992); L. Brey, J. J. Palacios, and C. Tejedor,
ibid {\bf 47},13884 (1993); T. Suzuki and Tsuneya Ando, J. Phys. Soc. Jpn. 
{\bf 62}, 2986 (1993).
\end{references}
\end{document}